\documentclass[aps,pra,twocolumn]{revtex4-1}

\usepackage{amssymb}
\usepackage{color,graphicx}
\usepackage{amsmath}
\usepackage{amsbsy}
\usepackage{amsthm}
\usepackage{bbm}
\usepackage{bm}
\usepackage{epsfig}
\usepackage{times}
\usepackage{float}
\usepackage{subfigure}
\usepackage[normalem]{ulem}

\newcommand{\Ket}[1]{\left\vert {#1} \right\rangle} 
\newcommand{\Bra}[1]{\left\langle {#1} \right\vert} 
 
\newcommand{\comm}[1]{}  
\newcommand{\proj}[1]{\left\vert {#1} \right\rangle \left\langle {#1} \right\vert} 

\newcommand{\ie}{{\it{i.e.}}}

\newcommand{\refeq}[1]{Eq.~(\ref{#1})}

\begin{document}

\title{Robust quantum state engineering through coherent localization in biased-coin quantum walks}
\author{H. Majury$^{1,2}$, J. Boutari$^3$, E. O'Sullivan$^{1}$, A. Ferraro$^{2}$, and M. Paternostro$^2$}
\affiliation{$^1$Centre for Secure Information Technologies (CSIT), QueenÕs University, Belfast BT7 1NN, United Kingdom\\
$^2$Centre for Theoretical Atomic, Molecular and Optical Physics,
School of Mathematics and Physics, Queen's University, Belfast BT7 1NN, United Kingdom\\
$^3$Clarendon Laboratory, University of Oxford, Parks Road, Oxford OX1 3PU, UK}
\date{\today}

\begin{abstract}
We address the performance of a coin-biased quantum walk as a generator for non-classical position states of the walker. We exploit a phenomenon of {\it coherent localization} in the position space --- resulting from the choice of small values of the coin parameter and assisted by post-selection --- to engineer large-size coherent superpositions of position states of the walker. The protocol that we design appears to be remarkably robust against both the actual value taken by the  coin parameter and strong dephasing-like noise acting on the spatial degree of freedom. We finally illustrate a possible linear-optics implementation of our proposal, suitable for both bulk and integrated-optics platforms. 
\end{abstract}
\maketitle



Quantum walks, which generalize the well-known concept of random walks to the quantum domain, have recently attracted considerable attention in light of the possibility that they offer the capability to explore non-trivial phenomena in statistical mechanics~\cite{kempe}, such as interference~\cite{jeong}, localization~\cite{crespi,crespi2,crespi3,crespi4} and tunnelling, and are able to establish entanglement among either the different degrees of freedom of a walker, or the parties of a multi-walker setting~\cite{ent,ent2,ent3,ent4,ent5,ent6,ent7}. Quantum walk-based architectures for quantum computation have been proposed and theoretically explored~\cite{childs}, while the possibility to implement quantum simulation through schemes based on the dynamics of a quantum walker are currently being pursued both theoretically and experimentally. 

In the last few years, the number of experimental validations of the quantum walk paradigm, and investigations towards its use for the coherent manipulation of information at the quantum mechanical level have flourished~\cite{refsJoelle,refsJoellea,refsJoelleb,refsJoellec,refsJoelled,refsJoelle2,refsJoelle2a,refsJoelle2b,refsJoelle2c,refsJoelle2d,refsJoelle3,refsJoelle3a,refsJoelle3b,refsJoelle4,refsJoelle4a}. This therefore paves the way to the full exploitation of the possibilities offered by quantum walks for quantum state engineering in Hilbert spaces of large dimensions, which is in general a difficult task to pursue experimentally. 

In this paper, we explore exactly such a possibility and exploit the statistical features of a discrete-time quantum walk to engineer non-classical states of an $N$-dimensional system. Specifically, we make use of special {\it coherent localization effects} induced on the position state of a quantum walker by choosing properly the operation describing the tossing of a coin. We demonstrate that high-quality coherent superpositions of fully position states of the walker can be arranged through this mechanism and simple post-processing operations. {As the focus of our investigation is on quantum state engineering, we consider explicitly the finite-$N$ case, and will not address explicitly the (undoubtedly relevant) scaling laws of the features that we discuss with the size of the walker space.} 

We also assess the robustness of our scheme against improper choices of the coin operation, showing that the scheme remains effective in a broad range of possible choices for such a transformation. Remarkably, such robustness extends to the effects of a dephasing environment that projects the state of the walker onto position states with no quantum coherence. Finally, we assess the experimental observability of the effects that we predict, showing that both bulk- and integrated-optics settings are suitable for such scope. 

The remainder of this paper is organised as follows: Sec.~\ref{Sec1} illustrates the basic features of the quantum walk at hand and present the effects induced by the use of a so-called Pauli $\hat Z$ operator. This will be sufficient to provide an intuition of the physical influences that low-parameter coin operations have on the statistical features of the walk. Sec.~\ref{Sec2} discusses the effects that deviations from the ideal conditions stated in Sec.~\ref{Sec1} have on the features of the walk, and how postselection aids in generating superposition states of the walker's position. Sec.~\ref{Sec3} discusses the case of a special preparation of the walker, and how this results in a quantum coherent superposition of spatial distributions, while Sec.~\ref{Sec4} illustrates a possible experimental implementation encompassing the features at the core of our analysis. Finally, Sec.~\ref{Sec5} presents our conclusion.

\section{Quantum Walk with SU(2) coin operation}
\label{Sec1}
The one-dimensional discrete-time quantum walk (DTQW) entails a `walker' endowed with two distinct degrees of freedom: a spatial one, which we dub `position', and an internal one, which embodies the `coin' that is tossed to decide the direction of the movement of the walker. The position state refers to where the walker is, in relation to the centre of the line on which it is moving. In the case of a linear walk and a dichotomic coin, the basis of the Hilbert space for such degrees of freedom are $\mathcal{H}_{p} :\{ \Ket{n}_p ; n \in \mathbb{Z}\}$ and $\mathcal{H}_{c} :\{ \Ket{0}_c, \Ket{1}_c\}$, respectively. 

Each step of the evolution comprises a coin-tossing operation and a position shift. 
The first can be formally described using the unitary coin transformation 
\begin{equation}\label{coinop}
\hat C_c(\theta) = \cos\theta \hat Z_c+\sin\theta\hat X_c
\end{equation}
with $\hat X,\hat Y,\hat Z$ the $x,y,z$ Pauli spin operator acting in ${\cal H}_c$. Although this is not the most general form of coin operation that can be devised, it captures the essence of the features that we aim at addressing in this work. Typically, the Hadamard coin $\hat C_c(\pi/4)$ is used, which yields equal probabilities for the walker to move leftward or rightward~\cite{su2}. The shift operator $\hat P_p$ acts on the position of the walker by displacing it conditionally on the state of the coin. We thus consider the joint operation  as
\begin{equation}
\begin{aligned}
\hat S&=\exp[i\hat P_p\otimes\hat Z_c]\\
&=\sum_{n}\left[\Ket{n+1}\Bra{n}_p\otimes\Ket{0}\Bra{0}_c+\Ket{n-1}\Bra{n}_p\otimes\Ket{1}\Bra{1}_c\right]
\end{aligned}
\end{equation}

We adopt a description based on density matrices right from this point, where $\hat\rho_{pc}(t)$ is the joint state of walker and coin at the discrete time $t$ of their evolution. With the notation introduced above we have 
\begin{equation}
\hat\rho_{pc}(t) = [\hat S ~\hat C_c(\theta)]\hat\rho_{pc}(t-1)[\hat S ~\hat C_c(\theta)]^\dag.
\end{equation}
Unless otherwise specified, it is intended that at the end of the walk the coin is traced out to leave room for an analysis of the statistical properties of the spatial degree of freedom. In this context, a key quantity in the study of quantum walk dynamics is the probability $P(n)={}_p\!\Bra{n}\text{Tr}_c[\hat\rho_{pc}(t)]\Ket{n}_p$ that position $n$ in the position space is occupied at the discrete time $t$. 

The one-dimensional DTQW with a Hadamard coin (dubbed here a Hadamard walk) has been the focus of extensive research activities both at the theoretical and experimental level~\cite{kempe,venegas}. The most striking feature of a Hadamard walk is the quadratic growth of the variance of the walk with the size of the lattice, a property that makes the DTQW very useful for the design of faster-than-classical search algorithms. 

On the contrary, walks with $\theta\neq\pi/4$ (thus resulting in unequal chances for the walker to move leftward and rightward)  have received only limited attention~\cite{su2}. Yet, interesting features emerge from considering biased coin operations. Most noticeably, for an $N$-step DTQW achieved using $\hat C_c(\theta)$, the variance of the walk has been shown to vary as~\cite{su2}
\begin{equation}
\sigma^2_\theta\sim(1-\sin\theta)N^2,
\label{sigma_theta}
\end{equation} 
which, for $\theta\in[0,\pi/4)$, is larger than that of a Hadamard walk. Notice that the initial state considered in Ref.~\cite{su2} is the factorized pure state  
\begin{equation}
\label{inistate}
 \hat\rho_{pc}(0) = \Ket{0}\Bra{0}_p \otimes \Ket{\phi}\Bra{\phi}_c \;,
\end{equation}
with $\Ket{\phi}_c=\frac{1}{\sqrt{2}} ( \Ket{0} + i\Ket{1})_c$.

Our goal is to further explore the features of a low-$\theta$ walk to see what properties of the walk depend on this choice of coin operation, and the extent of the control that can be operated on the walk through the tuning of such operation. The behaviour for low-$\theta$ described by \refeq{sigma_theta} can be intuitively understood considering the limit case of $\theta=0$, corresponding to a Pauli $\hat Z_c$ coin operation. For a generic initial pure state of the walker $\Ket{\psi_0}_p$, the state of the system after $t$ steps reads
\begin{equation}
\Ket{\Psi(t)}_{pc}=\frac{1}{\sqrt2}\left[
\Ket{\psi_{-t}}_p\Ket{0}_c + i (-1)^t \Ket{\psi_{+t}}_p\Ket{1}_c 
\right],
\label{zero_theta}
\end{equation} 
where the walker states $\Ket{\psi_{\pm t}}_p$ are defined as the initial state $\Ket{\psi_0}_p$ displaced by $\pm t$ steps in the position space $\mathcal{H}_{p}$
\begin{equation}
\Ket{\psi_{\pm t}}_p=\sum_n\Ket{n \pm t}_p \Bra{n}\psi_0\rangle_p.
\end{equation} 
If the initial state has support over a finite $M$ of sites, then the state in \refeq{zero_theta} is maximally entangled between the coin and the walker for any $t > \lceil M/2 \rceil$, regardless of the specific shape of the initial wave function. The state of the walker is rigidly displaced towards the extremes of $\mathcal{H}_{p}$.Therefore, if the initial state $\Ket{\psi_0}_p$ is localised in the position state, we have a coherent superposition of two states localized towards the opposite ends of the walker line. In addition, the absence of quantum interference in the position space explains the larger variance of the $\hat Z_c$-walk with respect to the Hadamard walk, as per \refeq{sigma_theta}. 

\section{Deviations from an ideal $\hat Z_p$-walk}
\label{Sec2}

Given the behaviour just described, a natural question arises: are the main features of the $\hat Z_c$-walk --- namely, the generation of maximal entanglement and large variance --- robust against imperfections? Here we will focus on two main sources of imperfections. First, we will consider the case in which the coin parameter $\theta$ is different from zero. Second, we will take into account the detrimental effect of noise on the walk. 

\subsection{Robustness to small $\theta$ deviations}
 
\begin{figure}[t!] 
\includegraphics[width=\columnwidth]{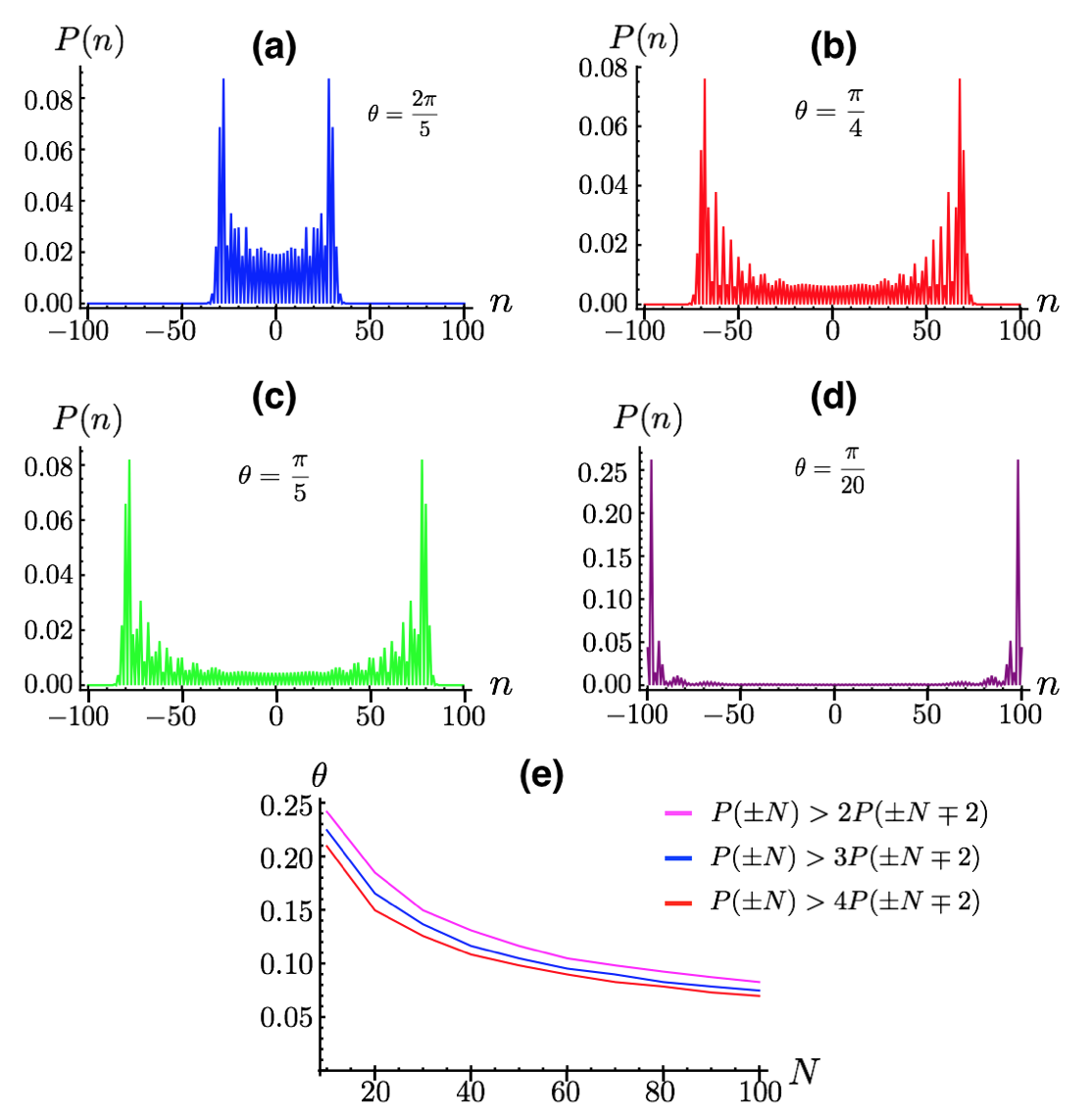}
\caption{(Color online) \textbf{(a)-(d):} Probability distributions for the DTQW with initial state $\Ket{0}_p\otimes\Ket{\phi}_c$ [with $\Ket{\phi}_c$ as in Eq.~\eqref{inistate}] for varying coin parameter $\theta$ and $N=100$ steps. As $\theta$ decreases, the walk spreads much wider on the lattice, thus enabling the walker to visit more sites on the line. For $\theta={\pi}/{20}$ the walker manages to spread to all the sites on the line, reaching the end-site on the end-lattice positions $\Ket{-100}_p$ and $\Ket{100}_p$.  Moreover, the probability to occupy sites different from the end ones is strongly reduced, thus manifesting a pronounced coherent localization effect of the walker. \textbf{(e):} Shown is the value of $\theta$ needed for the probability of the walker being at the end site of the lattice to be double (top line), triple (middle line) and four times (bottom line) the probability of being at the previous site. As the number of sites in the lattice increases, the critical value of $\theta$ needed to reach the end site decreases. Note that we compare the probability of being at end sites $\pm N$ with the probability of being at site $\pm N \mp 2$. This is due to the action of the shift operator, which allows us to fill only odd or even positions, depending on how many steps the walker takes. When the evolution occurs over an even number of steps, the walker can only occupy even sites on the lattice. Therefore, $P(N \pm 1)$ will always equal 0. 
}
\label{nonoise}
\end{figure}

We first look at the ideal scenario corresponding to the full absence of noise and arbitrary $\theta$, in order to gain insight into the phenomenology of the walk. Considering the initial state in Eq~(\ref{inistate}), Fig.~\ref{nonoise} shows the probability distribution $P(n)$ of finding the walker at the $n^\text{th}$ site of a lattice after $N=100$ steps when the coin parameter is increased.

\begin{figure}[b] 
{\bf (a)}\hskip3.5cm{\bf (b)}
\includegraphics[width=0.51\columnwidth]{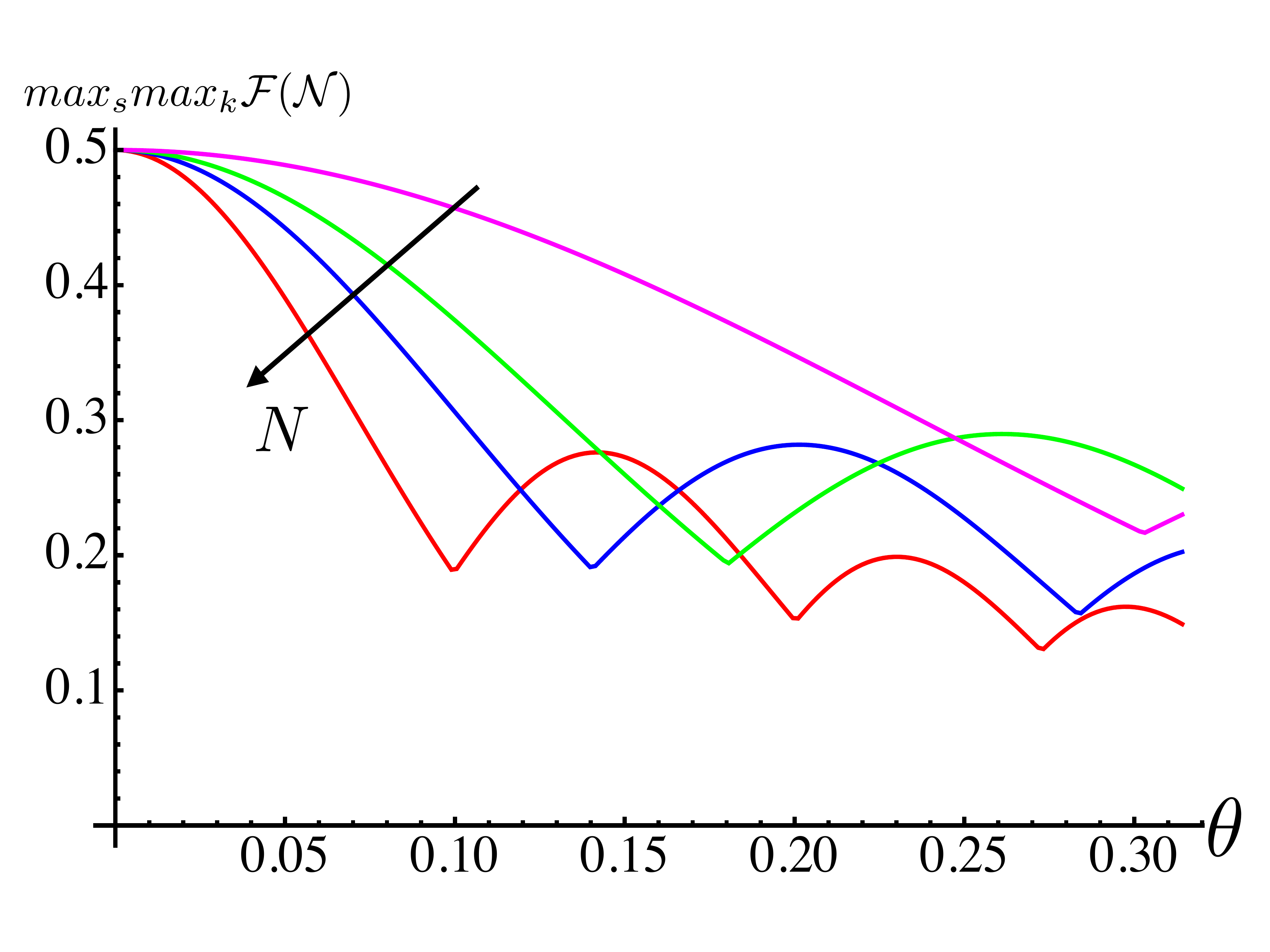}\includegraphics[width=0.51\columnwidth]{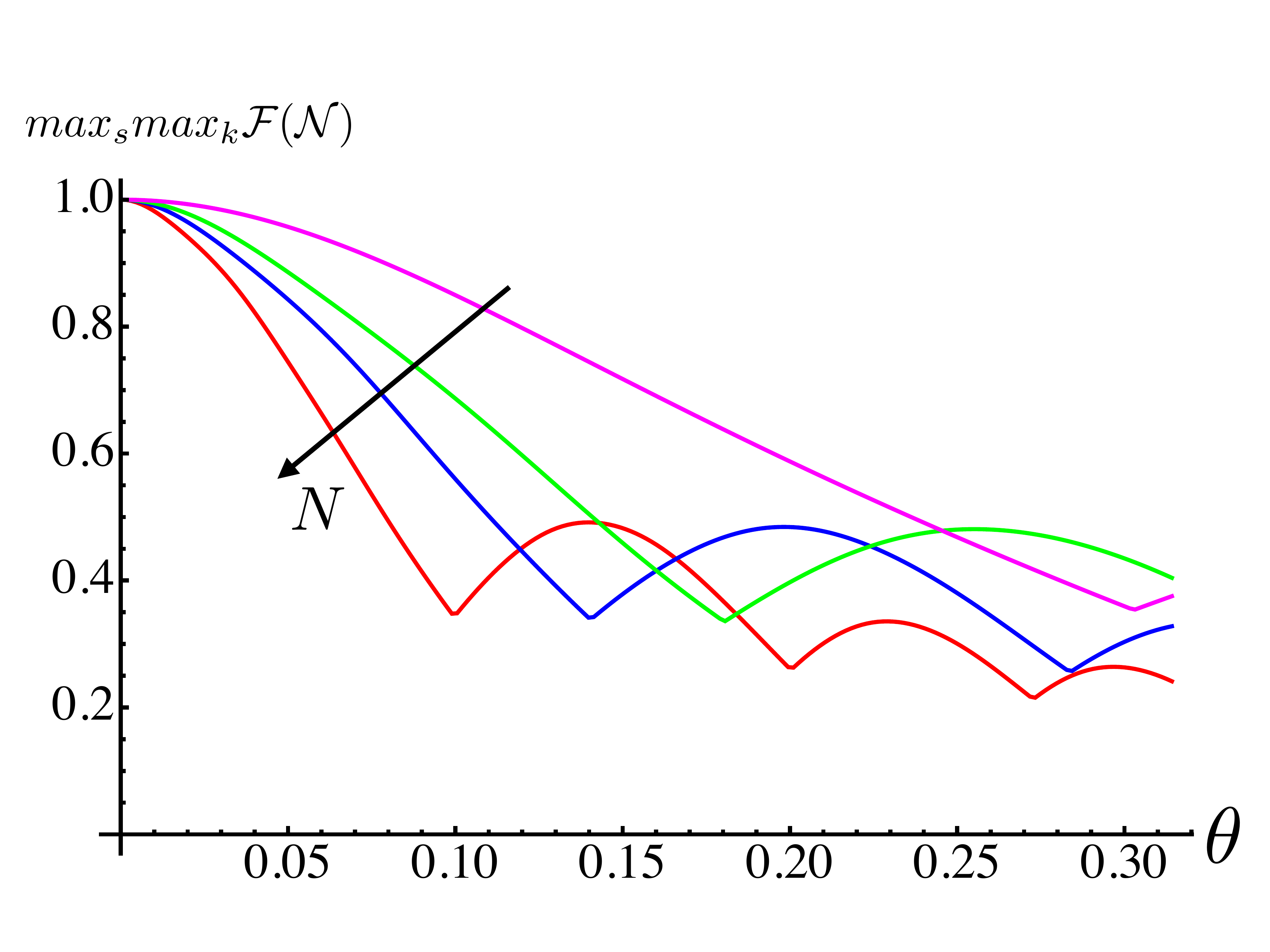}
\caption{{\bf (a)}  We plot the state fidelity $\max_{s}\max_{k}{\cal F}(N)$ evaluated over the target state $(\Ket{-N+k}_p +s \Ket{N-k}_p)/\sqrt2$ (with $s=\pm1$, $k=0,1,2,3$) and the position state of the walker after $N=10$ (magenta, top curve), $N=30$ (green, second curve), $N=50$ (blue, third curve) and $N=100$ (red, bottom curve) steps, plotted against the coin parameter $\theta$. The initial monotonically decreasing trait of each curve corresponds to the fidelity with the state with $k=0$. The first {\it ripple} is for $k=1$, and so on. {\bf (b)} Same as panel {\bf (a)} but for the conditional state of the walker achieved by projecting the coin onto its initial state. Despite the similarity between the curves shown in panel~{\bf (a)} and~{\bf (b)}, they are not related by a scaling factor: the ratio between one of the curves in panel {\bf (b)} and the corresponding one in panel {\bf (a)} is a non-constant function of both $N$ and $\theta$.}
\label{ControTheta}
\end{figure}

The effects of such variations on the probability distribution of the walk are significant: As $\theta$ increases from ${\pi}/{4}$, the spread of the walk decreases, while a decrease of $\theta$ from the value corresponding to a Hadamard walk results in the walker being able to spread faster and visit more sites of the lattice, in line with Ref.~\cite{su2}. Remarkably, for small values of the coin parameter, the walker manages to explore all the sites on the line, with the lobes of the probability distribution reaching the opposite ends of the position space [cf. Fig.~\ref{nonoise} {\bf (d)}]. Similarly to what is observed for the ideal $\hat Z_c$-walk, there is a strong modification of the interference mechanism responsible for the features of a quantum walk. This results in a significant reduction of the probability for the walker to occupy sites on the lattice that are different from those close to the end of the position space. A closer look reveals that a small value of the coin parameter effectively localizes the walker around $n=\pm N$ in a coherent way (the underlying dynamics of the walker is fully unitary, at this stage), although the probability for the walker to occupy {\it exactly} such end sites appears to depend on the size of the position space itself. 

In order to characterize this feature, we use the state fidelity~\cite{NC}
\begin{equation}
{\cal F}(N)={}_p\langle T(l)\vert\hat \rho_p(N)\vert T(l)\rangle_p
\end{equation}
between the reduced state of the walker $\hat\rho_p(N)={\rm Tr}_c[\hat\rho_{pc}(N)]$ and a chosen target state of the form $\vert{T(l)}\rangle_p=(\Ket{+l}_p+s\Ket{-l}_p)/\sqrt{2}$ (with $s=\pm1$ and $l\in{\mathbb N}$). As $\theta$ grows, we find that the fidelity with $\Ket{T(l)}$ improves by considering decreasing values of $l$. This is shown in Fig.~\ref{ControTheta} {\bf (a)}, where we have taken the target states 
 $\Ket{T(N-k)}_p$ (for $k=0,1,2,\dots$) and plotted $\max_{s}\max_{k}{\cal F}(N)$ against the value taken by the coin parameter and for growing sizes of the the position space. We consider these superposition states for comparison with the projected case and on the basis of the insight given by the analytical solution for $\theta=0$, see \refeq{zero_theta}. While the fidelity with the state with $k=0$ decreases as the coin parameter grows, the state of the position degree of freedom becomes close to coherent superposition states with growing values of $k$, as witnessed by the {\it ripples} displayed in Fig.~\ref{ControTheta} {\bf (a)}.  

\begin{figure*}[t!] 
{\bf (a)}\hskip4cm{\bf (b)}\hskip4cm{\bf (c)}\hskip4.0cm{\bf (d)}\\
\includegraphics[width=0.51\columnwidth]{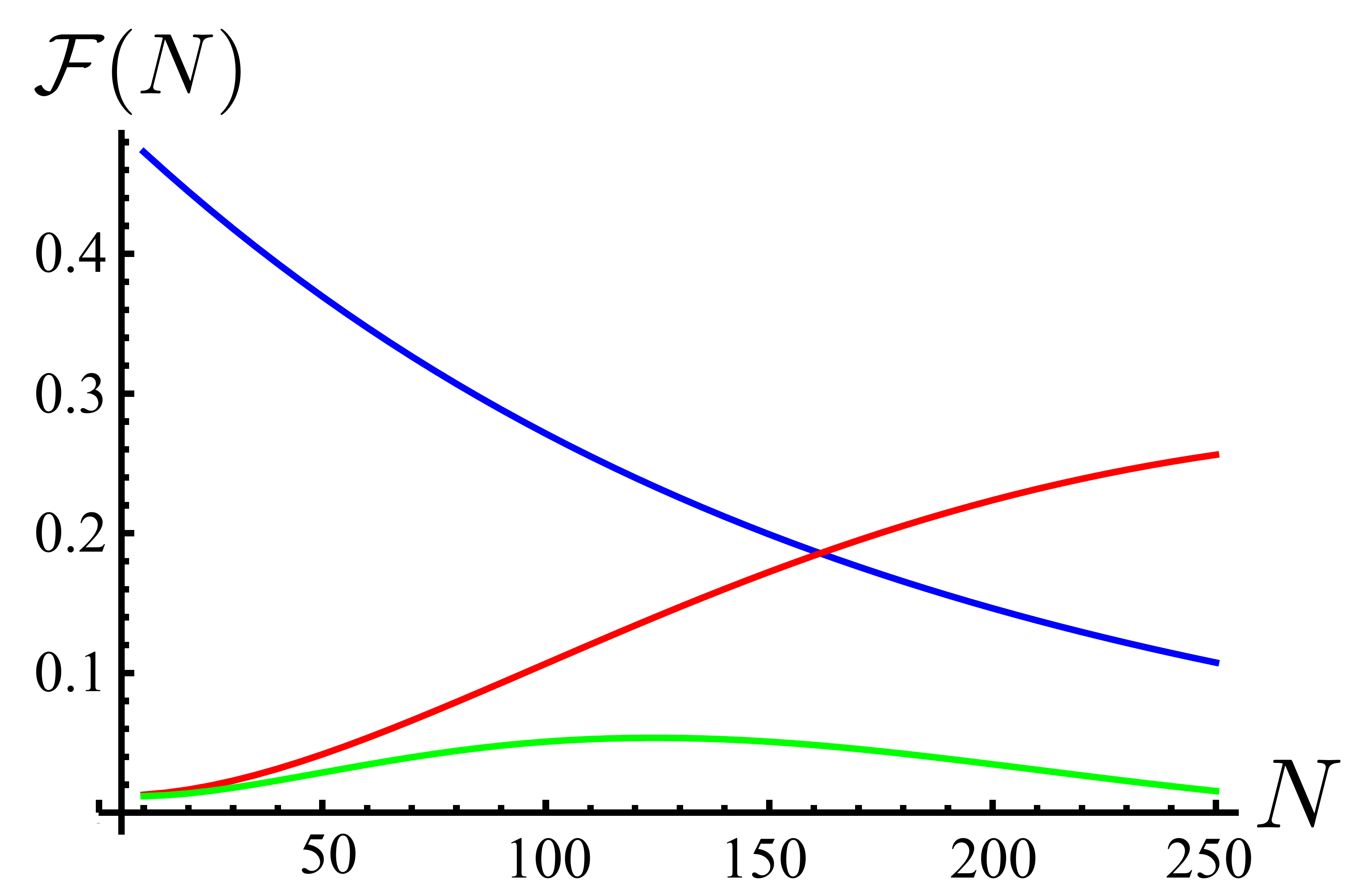}\includegraphics[width=0.51\columnwidth]{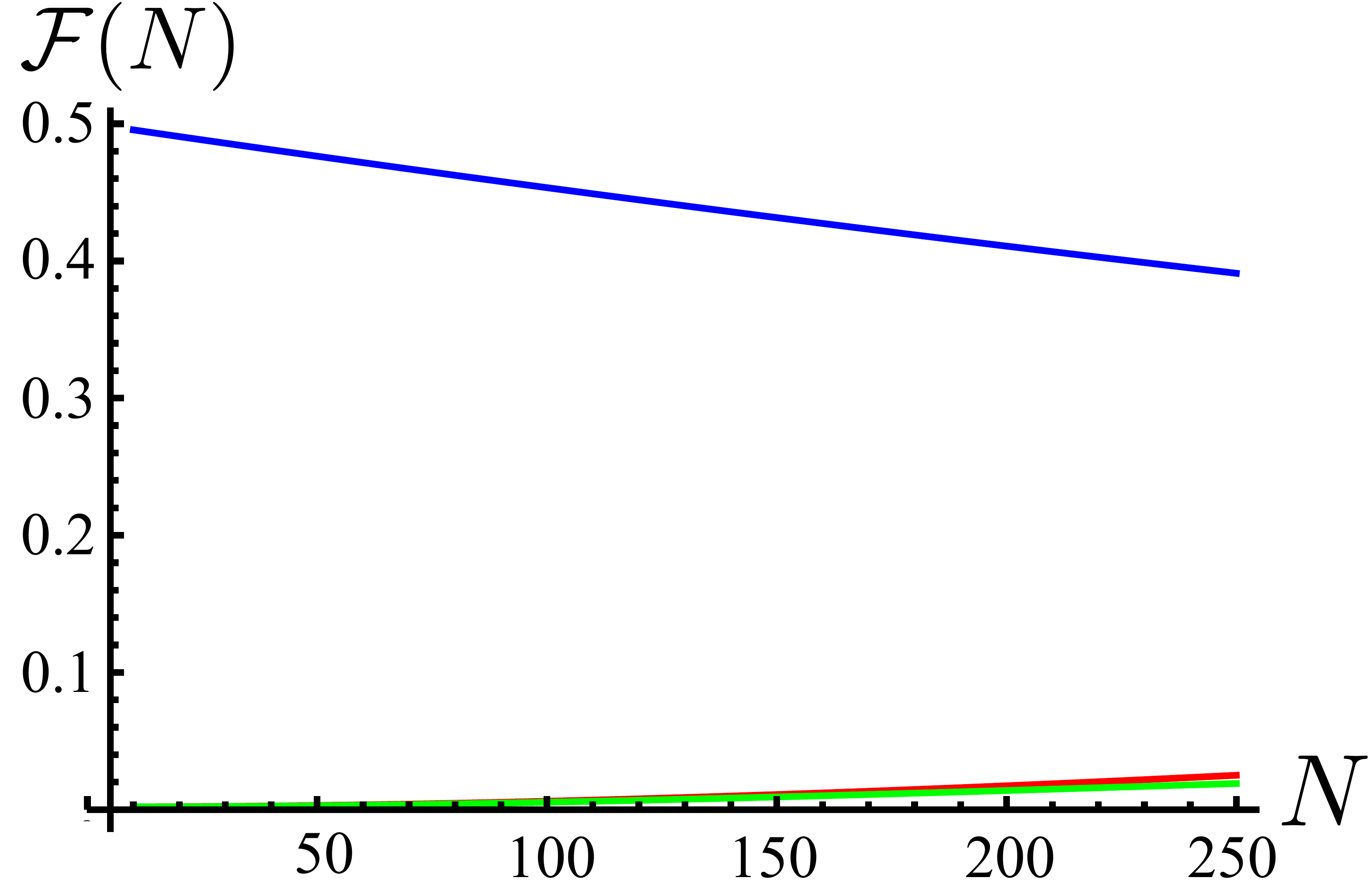}
\includegraphics[width=0.51\columnwidth]{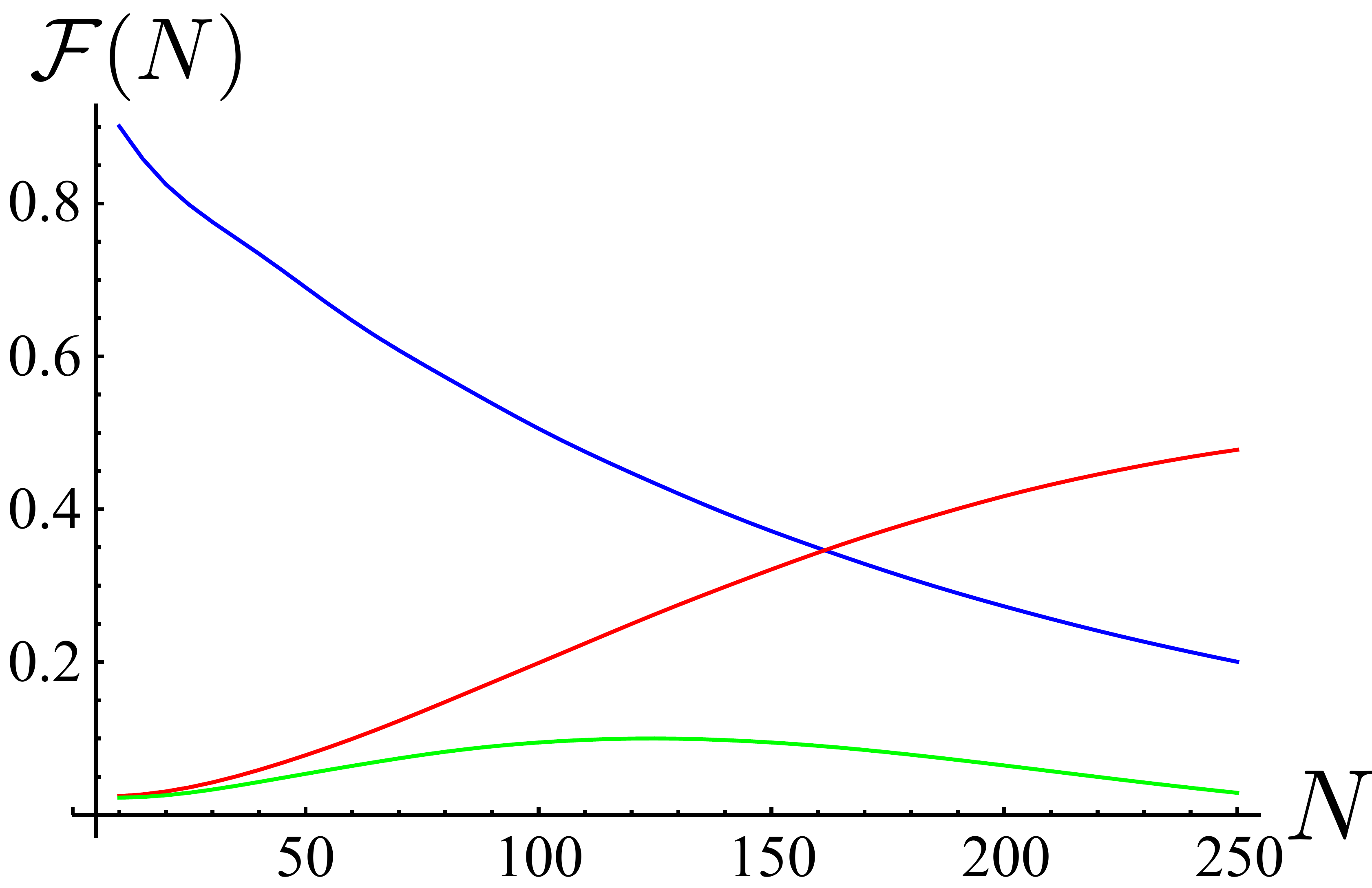}\includegraphics[width=0.51\columnwidth]{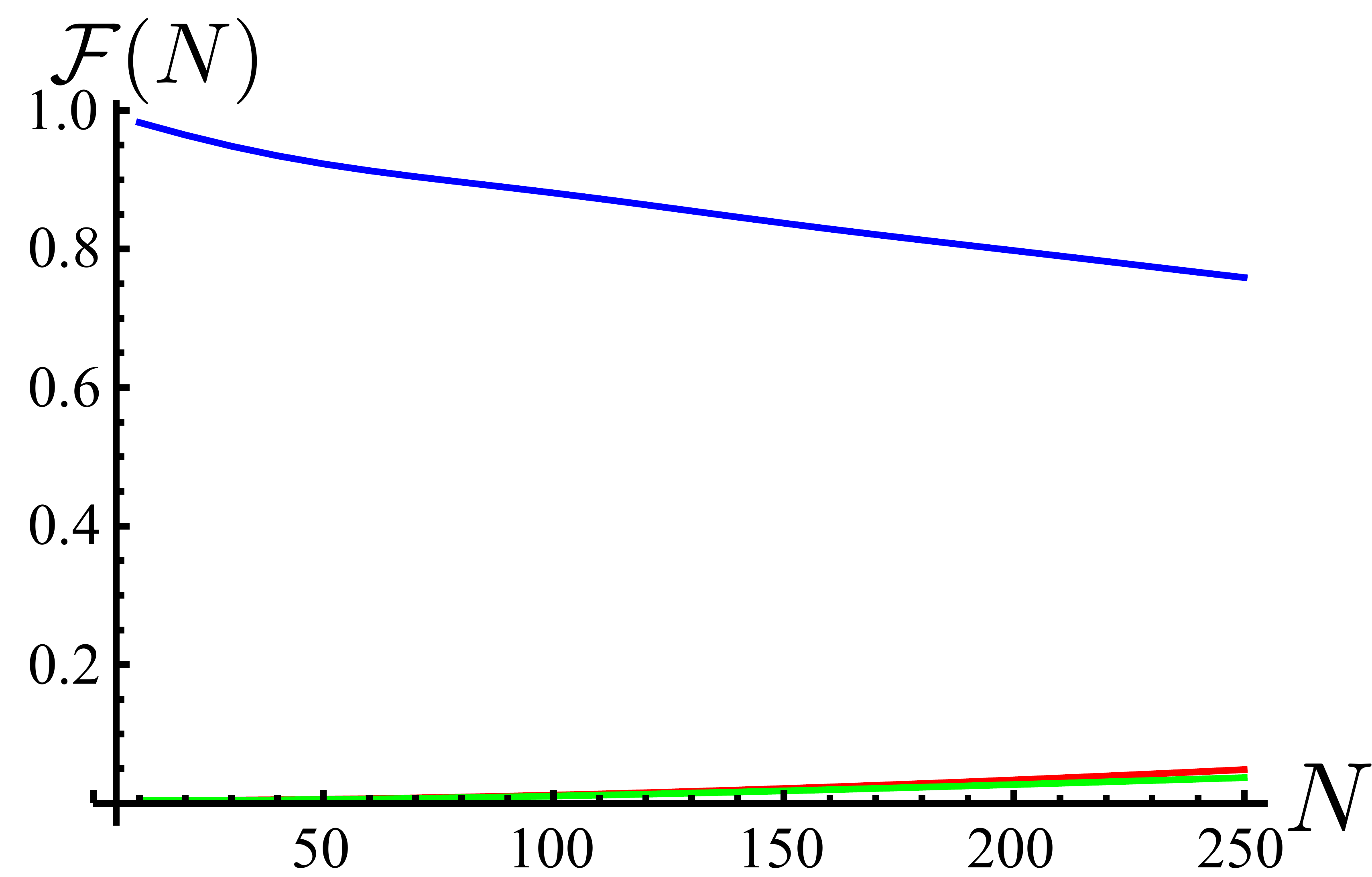}
\caption{{\bf (a)} State fidelity between the position state of the walker at the end of a $2N+1$-site walk (obtained by tracing away the coin's degrees of freedom) with $\theta=\pi/40$ (initial state $\Ket{0}_p\otimes\Ket{\phi}_c$) and a coherent superposition states of the form $(\Ket{-N+k}\pm\Ket{N-k})/\sqrt2$. We have considered the cases of $k=0$ (blue, top curve), $k=1$ (red, middle curve), and $k=2$ (green, bottom curve). {\bf (b)} Same as panel {\bf (a)} but for $\theta=\pi/100$. {\bf (c)} and {\bf (d)} Same analysis as in panels {\bf (a)} and {\bf (b)}, respectively, but for a conditional position state achieved by projecting the coin onto $\Ket{\phi}_c$ at the end of the walk. }
\label{cat}
\end{figure*}

Notice that, in the case of a $\hat Z_c$-walk, the walker-coin entanglement and the rigid translation of the initial state towards the extremes of the walker line implies a final walker state $\hat\rho_p(N)=\frac12(\proj{N}+\proj{-N})$. The latter has a fidelity $\mathcal{F}=1/2$ for $k=0$ independently of $N$ which explains the low fidelity of Fig.~\ref{ControTheta} {\bf (a)}. We can see that, in case of low $\theta$, the residual entanglement established between coin and position of the walker prevents the state of the walker from exhibiting strong coherences between $\Ket{N-k}$ and $\Ket{-N+k}$. The absence of such coherences is the factor limiting the values taken by the fidelity studied in Fig.~\ref{ControTheta} {\bf (a)}. The situation can be significantly modified if we slightly change our approach to the determination of the position state of the walker: rather than tracing out the coin after the $N$-step walk, we projectively measure the coin state in the basis determined by $\{ \Ket{\phi}_c, \Ket{\phi_\perp}_c\equiv\frac{1}{\sqrt{2}} ( i\Ket{0} +\Ket{1})_c\}$. That is, we consider the conditional position state
\begin{equation}
\hat\rho^j_p(N)=\frac{\text{Tr}_c\left[\hat\Pi^j_{c}\hat\rho_{pc}(N)\hat\Pi^j_c\right]}{\text{Tr}[\hat\Pi^j_{c}\hat\rho_{pc}(N)]},
\label{conditional}
\end{equation}
where $j=0,1$ and $\hat\Pi^0_c\equiv \proj{\phi}_c$ and $\hat\Pi^1_c\equiv\proj{\phi_\perp}_c$. Notice that in the case of a $\hat Z_c$-walk, one has
\begin{equation}
\begin{aligned}
\hat\rho^j_p(N)=\tau^j_p(N)&\equiv\frac12\left[
\Ket{-N}\Bra{-N}_p+\Ket{N}\Bra{N}_p\right.\\
&\left.+s^{N+j}(\Ket{N}\Bra{-N}_p+h.c.)
\right],
\label{target}
\end{aligned}
\end{equation}
which shows strong coherences in the position basis. Here $\tau^j_p(N)=\Ket{T(N)}\Bra{T(N)}_p$ is the density matrix of a walker prepared in a superposition over the end sites of the lattice. This feature is preserved for the case of low $\theta$, as demonstrated in Fig.~\ref{ControTheta} {\bf (b)}, where the fidelity $\max_{s}\max_{k}{\cal F}(N)$ with the target states $\tau^j_p(N-k)$(where $k= 1,2,3...$) is substantially larger than the corresponding one for the unconditional state over the whole range of lattice size that we have considered. The analysis at set size of the position space and changing $\theta$ presented above is completed by the complementary study against the size of the walker space showcased in Fig.~\ref{cat} {\bf (a)}-{\bf (d)}. Fig.~\ref{ControTheta} {\bf (b)}, Figs.~\ref{cat} {\bf (c)}-{\bf (d)}, and Eqs.~(\ref{conditional}) and (\ref{target}), showing that the high level of walker-coin entanglement determines coherent superpositions of localised states --- implying that the main features of the $\hat Z_c$-walk are retained also for small deviations of the coin parameter from $\theta=0$.


\subsection{Robustness to dephasing noise}

The striking differences between the probability distribution for small-$\theta$ DTQW and the one associated with a Hadamard walk persist under the influences of relevant forms of noise on the walk itself. In light of the transport-like nature of the DTQW mechanism, and the core role played by interference in the establishment of this salient feature, it is appropriate to focus on a dephasing-like channel $\Phi^t_{p}[\hat\rho_{pc}(t)]$ affecting the position degree of freedom only and that transforms the state $\hat\rho_{pc}(t)$ of the walker at time $t$ into 
\begin{figure}[b]
\includegraphics[width=1.01\columnwidth]{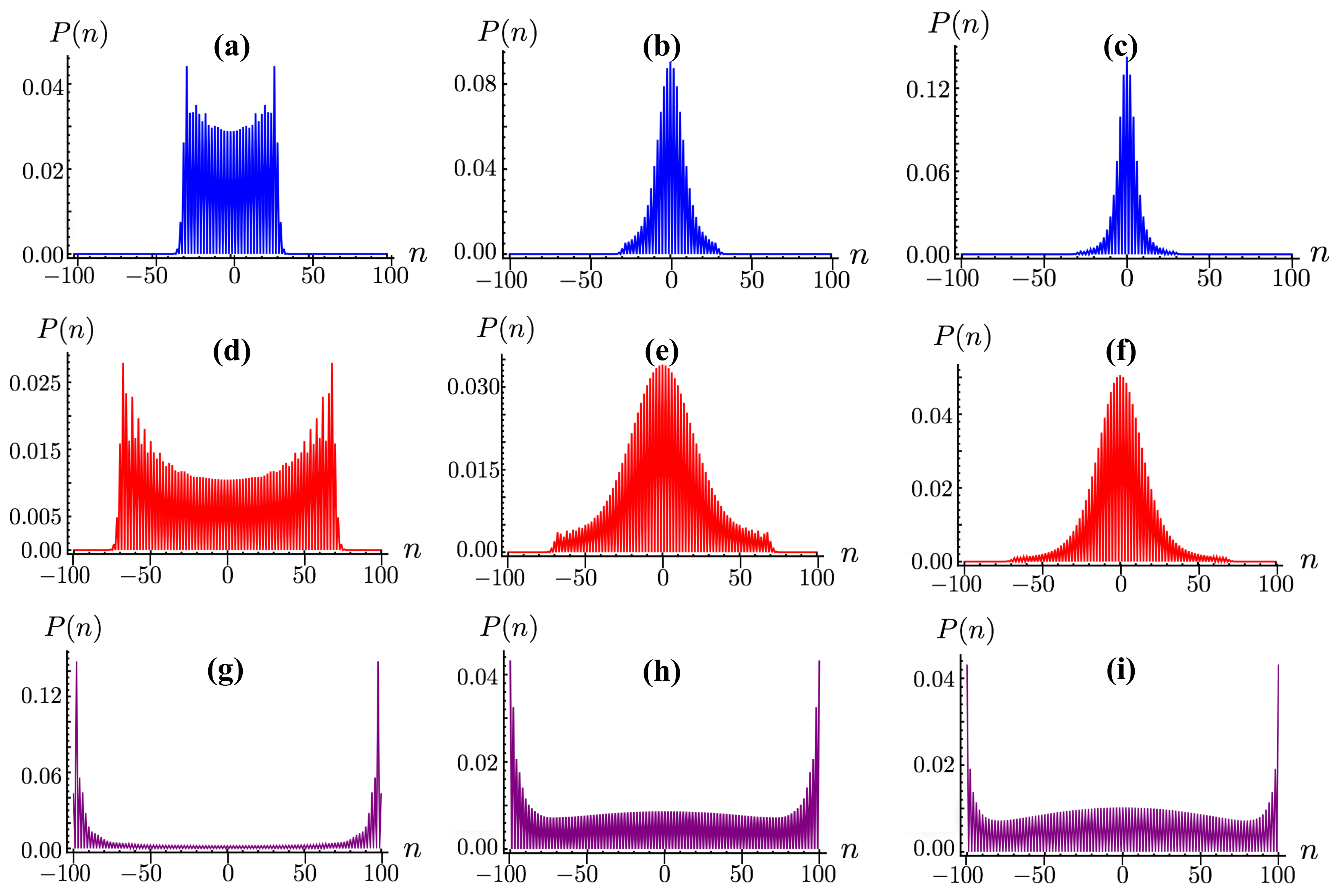}
\caption{(Color online) Panels {\bf (a)}-{\bf (c)}: Probability distributions for the DTQW with initial walker state $\Ket{0}_p \otimes \Ket{\phi}_c$ for $N=100$, $\theta=2\pi/5$ and a growing noise amplitude. We have taken $f = 0.05$ in panel {\bf (a)}, $f = 0.5$ in panel {\bf (b)}, and $f=1$ in panel {\bf (c)}. Panels {\bf (d)}-{\bf (f)}: Same as above but for $\theta=\pi/4$, thus implementing a Hadamard walk. Panels {\bf (g)}-{\bf (i)}: Same as above but for $\theta=\pi/20$. In each case we have obtained these results by averaging over 100 realizations of the noise-affected walk.
}
\label{pdf1}
\end{figure}
\begin{equation}
\hat \rho_{pc}(t)\to\Phi^t_{p}[\hat\rho_{pc}(t)]={\beta}_t \hat \rho_{pc}(t)  + (1-\beta_t) \!\!\!\sum^N_{k=-N} \Ket{k}\Bra{k}_p\otimes \hat\varrho^k_{c}(t),
\label{noise}
\end{equation}
where $\beta_t\in[\delta,1]$ is a discrete time-dependent noise parameter, whose value we chose randomly at every step of the walk, and $\hat\varrho^k_{c}(t)=\Bra{k}\hat\rho_{pc}(t) \Ket{k}$ are operators in ${\cal H}_c$. Here, $\delta\in[0,1]$ determines the range of probabilities within which $\beta_t$ is chosen at a given step of the evolution. We choose $\beta_t$ to be a random variable to ensure our theory and results are more consistent with what we would recover in an experimental setting. It is unlikely that there would be a constant, fixed value for the amount of noise within an experiment. Most likely it will fluctuate within a range, which we allow for by our choice of $\beta_t$. In what follows, we will refer to the amplitude $f=1-\delta$ of such range as the {\it noise amplitude}. Therefore, at step $t$ (with probability $\beta_t$) the mechanism that we consider leaves the state of the walker unaffected. With a complementary chance it renders it an incoherent admixture of position states, thus erasing any previously set spatial coherence. The dynamics of the walker thus proceeds as follows
\begin{equation}
\label{withnoise}
\hat\rho_{pc}(t)=\Phi^t_p[(\hat S ~\hat C_c(\theta))\hat\rho_{pc}(t-1)(\hat S ~\hat C_c(\theta))^\dag].
\end{equation}
By varying the range within which the values of the set $\{\beta_t\}$ are taken, we go from the ideal case with no noise, to the maximum noise interval with substantial dephasing effects. 

Previous investigations focused on the Hadamard walk have shown that, in the presence of maximum noise, the distribution $P(n)$ tends to the Gaussian distribution for a classical walk~\cite{noisepaper}. On the other hand, small amounts of noise have been shown to be beneficial for the spreading of the walker's distribution, thus demonstrating dephasing assisted-like walk~\cite{tregenna}. Here we aim at exploring the resilience of the localization effect observed at small values of $\theta$ when the walk is affected as in Eq.~\eqref{withnoise}. 

It is instructive to first consider analytically the behaviour of a $\hat Z_c$-walk in the presence of full dephasing noise, where $\beta_{t}$ is a constant rather than a random variable (\ie, $\theta=\beta_t=0$). Considering a generic initial pure state for the walker $\Ket{\psi_0}_p$, 
a straightforward calculation shows that after $t$ steps the state of the system is given by
\begin{equation}
\hat\rho_{pc}(t)=\frac{1}{2}\left[
\hat\psi_{-t}\otimes\proj{0}_c+\hat\psi_{t}\otimes\proj{1}_c
\right]\;.
\label{full_noise}
\end{equation}
In the equation above we have defined the following displaced states of the initial walker state, devoid of all its coherences
\begin{equation}
\hat\psi_{\pm t}\equiv \sum^N_{k=-N} |\Bra{k} \psi_0 \rangle |^2\Ket{k\pm t}\Bra{k \pm t}_p\;.
\end{equation}
\refeq{full_noise} is the incoherent counterpart of \refeq{zero_theta} and it shows that, in the presence of full dephasing noise, the system evolves towards an incoherent mixture of the two rigid translations of the walker initial state (devoid of all its coherences). If the initial state is localised in position, we thus have an incoherent mixture of two localised states. Evidently, this is at odds with the behaviour of the Hadamard walk in the presence of full noise, which is instead characterised by a Gaussian like distribution in position. The reason for this discrepancy is that, in the $\hat Z_c$-walk, the coin acts \textit{de facto} as a label that allows to rigidly translate the initial state of the walker towards the opposite ends of the position space, just as we have seen also for the ideal case of \refeq{zero_theta}. Both cases are then characterized by a localization effect at the extremes of the position space. 

Focussing now on the case of intermediate noise and arbitrary coin parameter $\theta$, Fig.~\ref{pdf1} summarises the findings of our investigation. For $\theta\ge\pi/4$, increasing noise strengths drive $P(n)$ towards a Gaussian-like distribution that is reminiscent of classical walks. In contrast with this, the localization resulting from the use of small coin parameters survives to arbitrary dephasing strengths, albeit enhancing the probability to find the walker at positions far from the end-lattice ones with respect to the $\hat Z_c$-walk (Fig.~\ref{pdf1} only shows a specific instance that does not affect the generality of our conclusions).

\section{Quenched spreading of DTQW with a small coin parameter}
\label{Sec3}
\begin{figure*}[t!]
\centering
{\bf (a)}\hskip5cm{\bf (b)}\hskip5cm{\bf (c)}\\
\includegraphics[width=0.65\columnwidth]{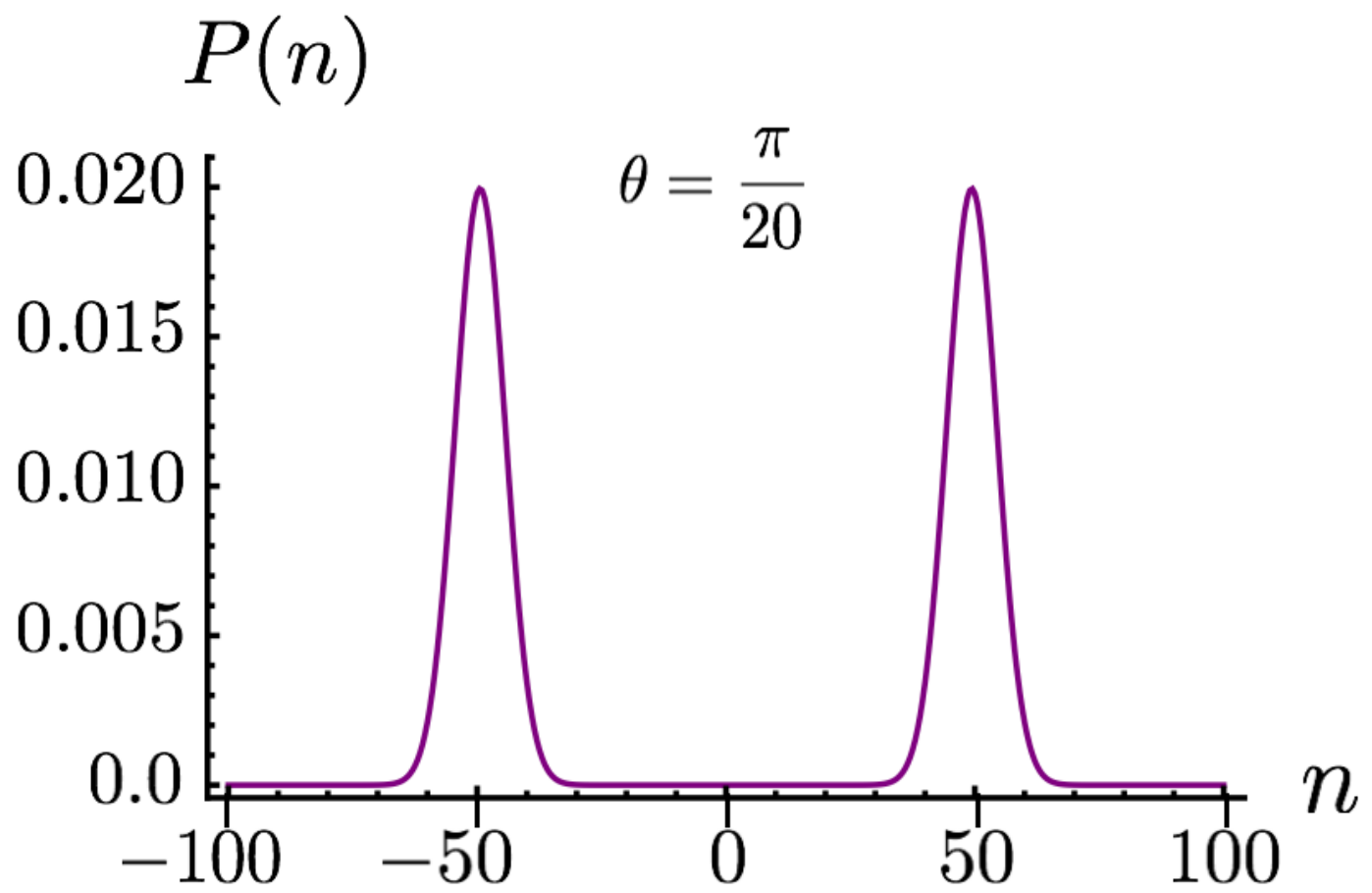}
\includegraphics[width=0.65\columnwidth]{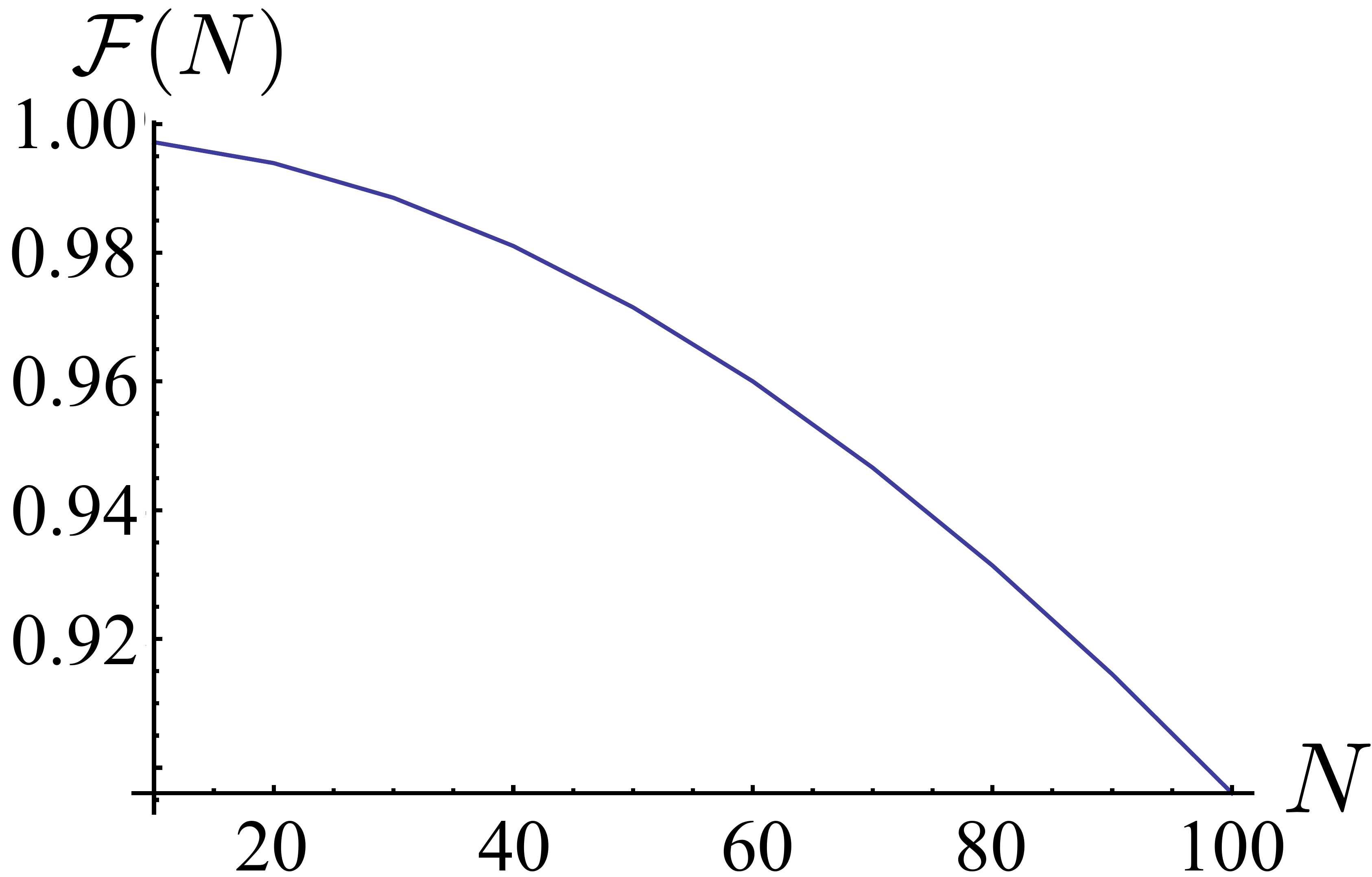}
\includegraphics[width=0.65\columnwidth]{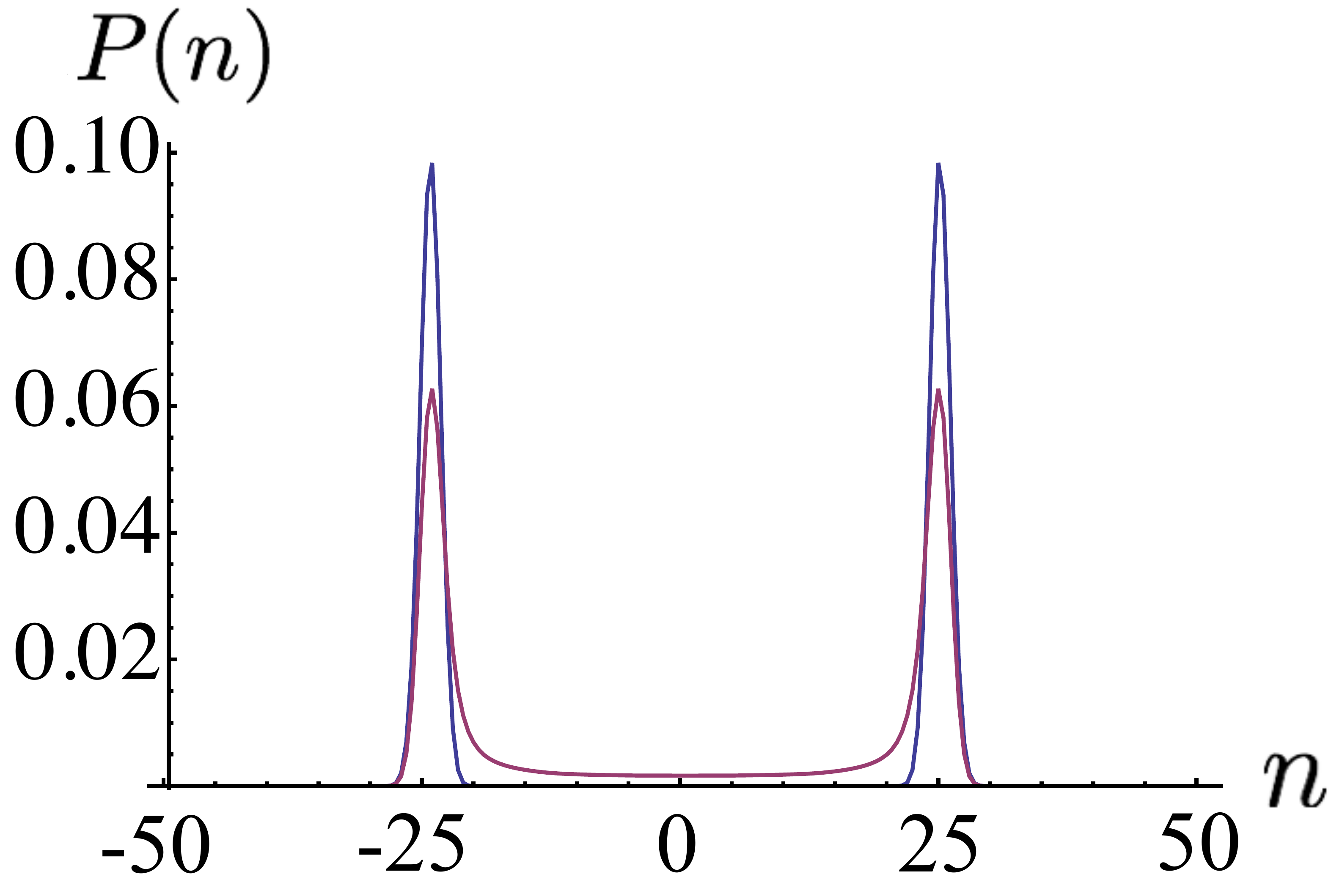}
\caption{(Color online) {\bf (a)} Probability distributions $P(n)$ against the site label $n$ for a DTQW with $N=50$ and the walker prepared in the initial state $\Ket{G}_p\otimes\Ket{\phi}_c$ with $\Sigma=10$. {\bf (b)} State fidelity between a Gaussian coherent quantum superposition state of the form in Eq.~\eqref{GaussCatState} and the conditional state of the position degree of freedom of a walker initially prepared in Eq.~\eqref{IniGauss} with $\Sigma=2$ and a growing size $N$ of the lattice. {\bf (c)} Comparison between the spatial probability distributions for a quantum walk with $\theta=\pi/20$ in the absence of noise (blue line, top peaks) and one achieved by taking $\{\beta_t\}\in[0.9,1]$ (magenta line, bottom peaks). The fidelity between the corresponding states is as low as 0.22.}
\label{GaussNoNoise}
\end{figure*}
The localization features discussed in the previous sections can be used for the sake of preparing interesting position states of the walker, or to preserve them from the natural dispersion entailed by the walk. Specifically, we first show that the small-coin parameter walk is able to {\it filter} high-quality coherent superposition states from specific initial position states of the walker. 

We start assessing the performance of a walker prepared in superpositions of spatial states. In this section we will see the quantum random walk whose initial position state is a superposition over the Gaussian and uniform distributions. For the Gaussian distribution, the initial position state is 
\begin{equation}
\label{IniGauss}
\Ket{G}_p = {\cal N}_G \sum_n e^{-\frac{n^{2}}{4\Sigma^{2}}}\Ket{n}_p,
\end{equation}
where $\mathcal{N_G}$ is the normalisation constant and $\Sigma$ is the standard deviation of the Gaussian. Fig.~\ref{GaussNoNoise} {\bf (a)} shows the probability distributions for $\Sigma=10$ and $\theta=\pi/20$, which is chosen here for convenience and clarity of presentation of the results. 

Quite clearly, two Gaussian peak appear, both of standard deviation $\Sigma$, separated roughly by $2N$. These features should be taken as canonical and valid for other choices of the coin parameter and lattice size. As for the case of a walker starting from the origin of the lattice, though, the distribution $P(n)$ thus found is actually associated with a state that is virtually deprived of coherences. A close inspection of the density matrix of the position degree of freedom only, in fact, reveals the virtual absence of off-diagonal elements. As done in the case of a walk starting from the origin of the lattice, we can enhance the quantum coherence in the position state of the system by projecting the coin onto $\Ket{\phi}_c$. This delivers a state that is very close to the quantum coherent superposition of spatial Gaussian distributions as
\begin{equation}
\label{GaussCatState}
\Ket{T}_p=\frac{1}{\sqrt2}\sum_n\left[{\cal N}_+e^{-\frac{(n-N/2)^2}{4\Sigma^2}}+{\cal N}_-e^{-\frac{(n+N/2)^2}{4\Sigma^2}}\right]\Ket{n}_p.
\end{equation}
Using again state fidelity as a figure of merit for the closeness of the walker's position state to $\Ket{T}_p$, we find the results reported in Fig.~\ref{GaussNoNoise} {\bf (b)}, which demonstrate the very high quality of the conditional state within quite a large range of lattice sizes. Fidelity remains above $92\%$ for $N\simeq80$, a result that appears to be only very weakly affected by an increase in the variance of the initial Gaussian distribution. 

On the other hand, the addition of dephasing noise strongly affects the features of the conditional position states. In Fig. 5 (c) we see that already for $\{\beta_t\}\in[0.9,1]$, the fidelity between the noiseless conditional state and the corresponding noise-affected one drops to $\simeq0.22$. 

\section{Experimental Proposal}
\label{Sec4}

\begin{figure}[b]
		\centering
		\includegraphics[width=0.5\textwidth]{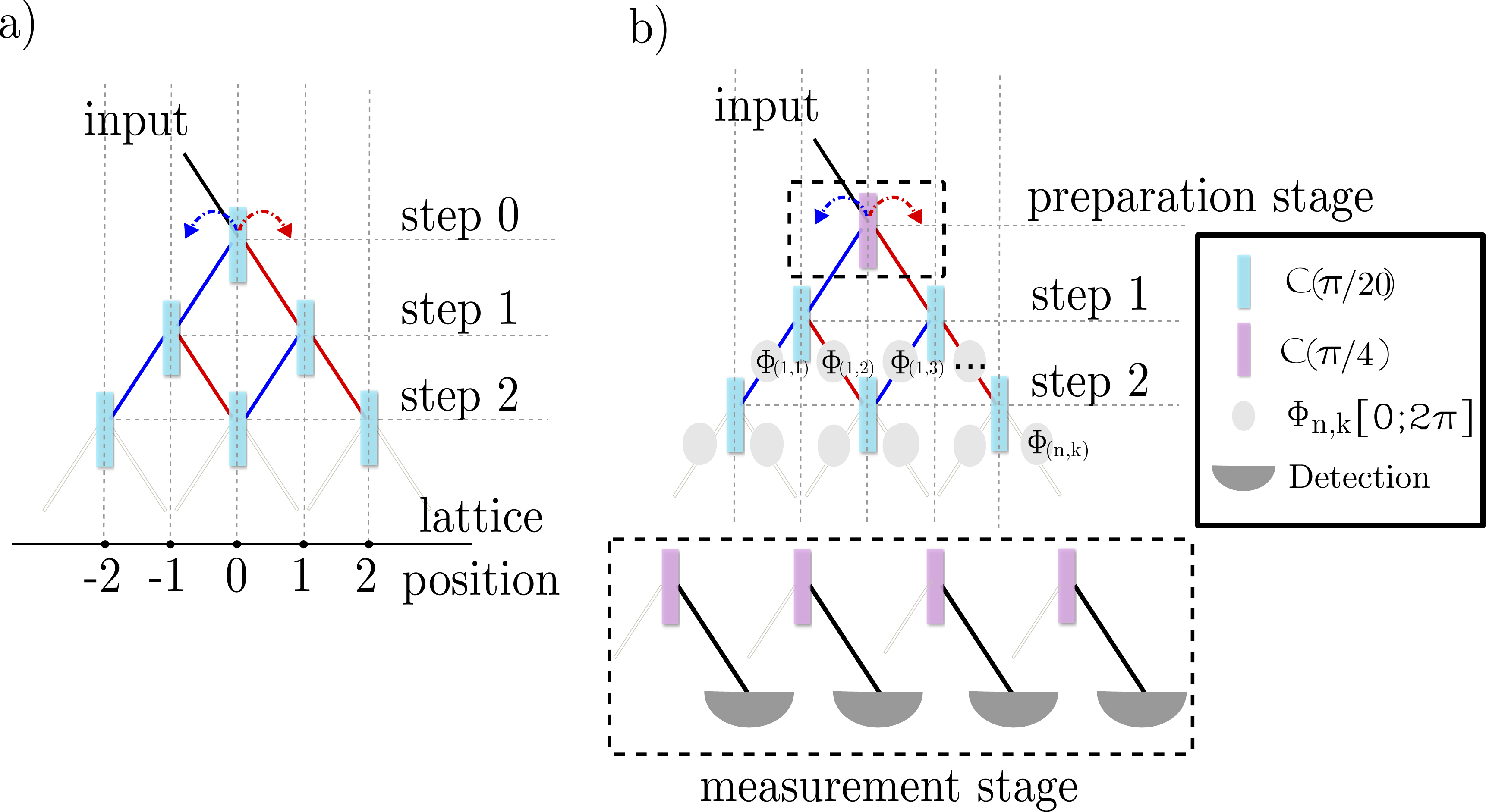}
\caption{(Color online) {\bf (a)} Beams propagating through a regular grid of beam splitters achieve a walk with spatial encoding. The powers of beams emerging from a grid of depth $N$ correspond to the probability distribution of the walker and coin (encoded in beam location and direction, respectively) after a walk of $N$ steps. {\bf (b)} Optical scheme for the generation of robust coherent superpositions of position states of the walker. The scheme consists of the preparation of the initial coin state $|\phi \rangle_c = \frac{1}{\sqrt{2}} (|0\rangle+i|1\rangle)$ through a beam splitter at $\theta = \pi/4$ where the output arms are then recombined to allow the walker to start the walk from position 0. The evolution is then performed with a biased coin, beam splitters with transmission ratio $\theta$, and a detection stage. The latter projects the coin onto a desired state through an array of beam splitters at $\theta = \pi/4$ and collecting the output from one side of arms. The phases $\phi_{n,k}$ where $n$ is the step and $k$ the sub-step index, indicate how noise can be inserted in the system.} 
		\label{Fig_exp}
\end{figure}

Here we present a brief description of a set up that can be used to implement a quantum walk scheme able to generate the sort of coherent state superpositions addressed in our analysis so far. 
In Fig.~\ref{Fig_exp} {\bf (a)} we show the optical spatial representation for a 1D DTQW. In this implementation, the coin is embedded into the walker's degree of freedom, which is embodied by the position of a light pulse, going through a multi-layer interferometer, on a detection line. The walker performs the first step by impinging on a beam splitter with a set reflectivity $\theta$. This effectively implements the coin-tossing operation, and prepares the coin state. The light beam is thus split into a superposition of directions, conditionally on such a coin operation. 

In order to reproduce the DTQW evolution with initial state $\hat\rho_{pc}(0)$ in \refeq{inistate}, we propose the following approach: a first coin operation is implemented to generate the state $|\phi \rangle_c$. This can be done through a $50:50$ beam splitter (i.e. we set the reflectivity ratio $\theta = \pi/4$). The outputs are then recombined to enter the first beam splitter of the walk [step $0$ in Fig.~\ref{Fig_exp} {\bf (a)}]. 
This is followed by the DTQW optical scheme, as in Fig.~\ref{Fig_exp} {\bf (a)}, where the beam splitters implementing the necessary series of coin tossing operations have a set reflectivity ratio of $\theta\ll1$. The process continues up to step $N$ [cf. Fig.~\ref{Fig_exp} {\bf (b)}]. In order to project on the initial coin state, after the final step, we project over the initial coin state inserting another row of beam splitters with $\theta = \pi/4$ and detecting on the output corresponding to $\frac{1}{\sqrt 2}(|0\rangle+i|1\rangle)$. 

In this context, an experimental test of the resilience of the localization phenomenon highlighted in this work can be performed by simulating dephasing noise on the position degree of freedom of the walk, which can be effectively implemented by inserting random phases $\phi_{n,k}$ between the beam splitters of consecutive walk steps. Here $n$ is the step index and $k$ the index of the modes into which the walker is split at a given step of the walk. The strength of such noise will be related to the rate with which we pick each phase between values $(0;2\pi]$ instead of choosing $0$.

A possible platform of the implementation of this scheme is provided by integrated optical waveguide  technology~\cite{crespi,crespi2,crespi3,crespi4}, where beam splitting operations are realised by optically {\it writing} on a chip a set of mode couplers, whose mixing ratio (equivalent to our coin parameter), can be chosen during the writing process. The main limitation of this architecture is the strength of losses experienced by a photonic signal propagating through. However, Ref.~\cite{Flamini} shows that integrated photonic architectures affected by standard fabrication and and propagation losses would allow for as many steps of a walk-like dynamics as 32 with only insignificant effects on the quality of the output walker state. The degree of reconfigurability of the large integrated-waveguides arrays is currently very limited, which would put some constraints to the possibility of simulating the effects of noise. 

A potential way around such limitations is provided by fiber-integrated setups exploiting time-multiplexing approaches. In this configuration, the encoding of the walker is done exploiting the temporal degree of freedom and the coin using polarisation or path degrees of freedom of optical signals. The conditional shift of the walker over the position space can be mapped into earlier and later arrival time given by two different path lengths~\cite{refsJoelle4,refsJoelle4a}. This might be realized using coupled loop cavities of different radius, such as in Ref.~\cite{joelle}. 
This approach would also be suitable for a dynamically variable phase to simulate noise in the system, enjoys more flexibility than integrated-waveguides settings, offering a potentially fruitful ground for the simulation of the noise effects addressed here, where averaging over many noise patterns would be needed. In order to project over the initial coin state either the last coin operation can be dynamically tuned back to $\theta = \pi/4$ via an electro-optical modulator (EOM) or the outputs can interfere over a beam splitter with $\theta = \pi/4$. The output will consist of a sequence of pulses in time-bins given by the size of the loops; the detection can be achieved through a single-photon avalanche diode with a time-to-digital converter that records the detection time relative to the initial pulse generation. The low losses in such a fibre implementation (Ref.~\cite{joelle} reports a high-fidelity quantum walk of up to 62 steps) would allow for the preparation of a superposition state of the form at the core of our study after 20-30 steps evolution. 

Finally, we point out that Ref.~\cite{LucaHelena} provides an exact, analytical methodology for the preparation of arbitrary states of a quantum walker through only passive linear-optics transformation. This would thus provide the way to engineer initial states of the walker such as the one in Eq.~(\ref{IniGauss}).
\\
\section{Conclusions}
\label{Sec5}

We have highlighted a phenomenon of coherent localization of the position of a quantum walker in its position space taking place when the parameter of the coin operation is chosen to be small. Such effect, which is a consequence of the enhanced variance of the quantum walk distribution resulting from the use of a small coin parameter~\cite{su2} turns out to be a remarkably efficient way to postselect important types of non-classical states. Our scheme offers features of robustness against the range of choices of the coin parameters, and the dephasing noise that might affect the position degree of freedom, while being suitable for agile experimental implementations in state-of-the-art linear optics experiments. Our study paves the way to the investigation of quantum state engineering based on the exploitation of the statistics of quantum walk processes, and reinforces the versatility of such mechanisms for low-control coherent operations on systems spanning naturally large Hilbert spaces. 

\noindent
{\it Note added} -- While completing this work, we became aware of the work by W.-W. Zhang, {\it et al.},   where  the absence of dispersion for delocalised initial states is used to engineer coherent superpositions of position states ~\cite{catstate}.

\acknowledgements 

We acknowledge support from the EU project TherMiQ, the Marie Curie ITN PICQUE, the John Templeton Foundation (grant number 43467), the Julian Schwinger Foundation (grant number JSF-14-7-0000), the UK EPSRC (grants EP/M003019/1, and EP/N508664/1), and the Northern Ireland DfE.  MP acknowledges the DfE-SFI Investigator Programme (grant 15/IA/2864).

\section*{Competing Interests}
The authors declare that they have no competing interests.

\section*{Authors Contribution}
AF and MP conceived the project with strong input from HM. HM performed the calculations and wrote the first draft of the manuscript. JB contributed with the design of the experimental proposal. All authors contributed to the finalisation of the manuscript.

\end{document}